\documentclass[conference]{IEEEtran}
\IEEEoverridecommandlockouts
\usepackage{cite}
\usepackage{amsmath,amssymb,amsfonts}
\usepackage{algorithmic}
\usepackage{graphicx}
\usepackage{textcomp}
\usepackage{xcolor}
\usepackage{url}
\usepackage{array}
\usepackage{enumitem}
\usepackage{listings}
\usepackage{tcolorbox}
\tcbuselibrary{skins,breakable}
\usepackage{algorithm}
\usepackage{algorithmic}

\newcommand{\promptfontsize}{\footnotesize}

\newcommand{\placeholder}[1]{%
  \textcolor{blue!70!black}{\texttt{\{#1\}}}%
}

\newtcolorbox{promptbox}[1][]{
  breakable,
  enhanced,
  colback=gray!5,
  colframe=gray!40,
  fontupper=\ttfamily\promptfontsize,   
  fonttitle=\bfseries\promptfontsize,
  title=#1,
  left=1em, right=1em, top=1em, bottom=1em,
  sharp corners
}

\begin{document}

\title{LLM-Empowered Functional Safety and Security by Design in Automotive Systems \\
\thanks{
This work has received funding from the European Chips Joint Undertaking under Framework Partnership Agreement No 101139789 (HAL4SDV) including the national funding from the Federal Ministry of Research, Technology and Space of Germany under grant number 16MEE00471K. The responsibility for the content of this publication lies with the authors.
}
}

\author{
\IEEEauthorblockN{
    Nenad Petrovic\IEEEauthorrefmark{1},
    Vahid Zolfaghari\IEEEauthorrefmark{1},
    Fengjunjie Pan\IEEEauthorrefmark{1},
    Alois Knoll\IEEEauthorrefmark{1}
}

\IEEEauthorblockA{\IEEEauthorrefmark{1}
    \textit{Chair of Robotics, Artificial Intelligence and Real-Time Systems} \\
    Technical University of Munich, Munich, Germany \\
    Email: \{nenad.petrovic, v.zolfaghari, f.pan, k\}@tum.de
}

}

\maketitle

\begin{abstract}
This paper presents LLM-empowered workflow to support Software Defined Vehicle (SDV) software development, covering the aspects of security-aware system topology design, as well as event-driven decision-making code analysis. For code analysis we adopt event chains model which provides formal foundations to systematic validation of functional safety, taking into account the semantic validity of messages exchanged between key components, including both CAN and Vehicle Signal Specification (VSS). Analysis of security aspects for topology relies on synergy with Model-Driven Engineering (MDE) approach and Object Constraint Language (OCL) rules. Both locally deployable and proprietary solution are taken into account for evaluation within Advanced Driver-Assistance Systems (ADAS)-related scenarios. 
\end{abstract}


\section{Introduction}
Rapid prototyping, verification, and deployment of advanced vehicle functions while preserving functional safety and cybersecurity is of utmost importance when it comes to innovation in automotive, especially for Software Defined Vehicles (SDVs) which are becoming more and more complex. Conventional workflows—relying on manually crafted specifications, state machines, and specialized modeling tools—offer the traceability required for compliance with ISO 26262 and ISO 21434. However, these processes require significant expert intervention and become increasingly difficult to scale as system complexity grows, especially when integrating diverse sensor modalities, distributed ECUs, and variable timing behaviors \cite{petrovic2025survey}.

Large Language Models (LLMs) offer new opportunities to accelerate these processes by extracting structure from natural-language requirements and generating preliminary behavioral models. Yet, when used naively, LLMs lack guarantees on correctness, fail-safe behavior, timing determinism, and security posture. Direct LLM-generated logic may introduce hallucinated signals, violate architectural interfaces, or ignore causal constraints critical for hazard analysis (HARA), failure propagation modeling, or threat analysis and risk assessment (TARA).

To address these gaps, we propose an event-chain–based, LLM-guided workflow designed explicitly to support functional-safety and security analyses. Event chains provide a structured representation of causal, temporal, and data-flow relationships that underlie safety mechanisms and timing requirements in automotive systems. By embedding event chains at the center of the LLM workflow, we enforce constraints necessary for safety argumentation: allowed signal flow, sequencing of safety-relevant steps, end-to-end latency budgets, and the separation of safety-critical and non-critical elements \cite{inc2021} \cite{inc2022} \cite{dingler2025}. Event chains thus act as formalized anchors that constrain LLM outputs and prevent unsafe or unverified behaviors from entering the generated artifacts.

Within this framework, the LLM assists in automating tasks normally requiring expert domain knowledge: deriving candidate safety mechanisms from textual requirements, aligning functional intent with the Vehicle Signal Specification (VSS), identifying potential hazards or unsafe interactions, and generating draft diagnostic or mitigation mechanisms. Retrieval-Augmented Generation (RAG) further limits the LLM’s reasoning scope to trusted data sources such as approved VSS catalogs, safety manuals, and architectural descriptions, reducing the likelihood of hallucinated signals or unassessed data paths. This ensures that automatically produced models maintain traceability to the authoritative sources required for safety and cybersecurity audits.

The synergy between event-chain models and LLMs enables automated pre-analysis of both functional safety (e.g., violation of time budgets, broken causal chains, missing safety reactions) and cybersecurity (e.g., unauthorized signal pathways, inconsistent trust boundaries, or unexpected control influence). Before any code is deployed to the target platform, the workflow performs consistency checks aligned with ISO 26262 Part 6 (software architectural constraints) \cite{iso26262} and ISO 21434 (secure signal and interface handling) \cite{iso21434}. As a result, LLM-generated artifacts not only become more reliable but also integrate directly into established safety engineering workflows.

Overall, the proposed pipeline contributes to idea of automating time-consuming automotive code safety and security analysis procedures, It aims to reduce the effort of early-phase design while preserving the rigor needed for modern safety and cybersecurity requirements to be satisfied. As a demonstration, we apply the workflow to ADAS scenarios based on our physical testbench \cite{wu2025viltum} and simulation platforms.

\section{Background and Related Works}
Large Language Models (LLMs) are increasingly incorporated into the development lifecycle of autonomous vehicle systems to tackle both safety and security concerns. From a safety perspective, models such as GPT-4 and CodeLlama are utilized for code co-generation in functions including Adaptive Cruise Control (ACC), and are assessed through automated evaluation pipelines employing Software-in-the-Loop (SIL) environments integrated with simulation to enable rapid preliminary validation. On the security front, dedicated frameworks such as the Security Test Automation Framework (STAF) integrate LLMs—such as GPT-4.1 and DeepSeek-V3—within self-corrective Retrieval-Augmented Generation (RAG) architectures to automatically derive executable security test cases from attack trees. Furthermore, LLMs like OpenAI’s Codex are used to generate formal security assertions, including SystemVerilog Assertions (SVAs), for low-level AV hardware components such as traffic signal controllers and AES encryption modules, with notably higher success rates observed when comprehensive design context is provided. In contrast, customized models like HackerGPT demonstrate the capability to automate the generation of cyberattack scripts targeting virtual Controller Area Network (CAN), Bluetooth interfaces, and key fob systems, highlighting the dual-use nature of LLMs and underscoring the necessity for robust and proactive automotive security measures. 

On the other side, in this paper we focus on security and safety by design, which aims to prevent malicious behavior of ADAS-related code before any execution on the target platform. In cases when threat or unsafe behavior risk are detected, our workflow also proposes corrective action in order to eliminate it. This way, our goal is to minimize the potential risk of damage to the environment caused by malicious code. It builds upon good practices observed within several of our previous works. In \cite{petrovic2025genai}, we presented integrated toolchain which generates ADAS simulation code in Python starting from textual requirements relying on LLMs in synergy with model-driven approach using Eclipse Modeling Framework (EMF), Ecore and Object Constraint Language (OCL). The focus was on vehicle configuration (sensors and actuators) and environmental aspects (placements of vehicle, pedestrian and obstacles). Additional details regarding the implementation of individual MDE-related steps can be found in our other papers: metamodelling \cite{petrovic2025meta}, instance model \cite{pan2025instance} and constraint generation \cite{pan2024ocl}. On the other side, in \cite{petrovic2025ec}, a workflow for LLM-aided C++ code generation aiming SDV zone ECU within physical testbench was presented, making use of event chain for vehicle's behavior validation. Both solutions rely on existing n8n \cite{n8n2025} workflow framework for toolchain integration. Additionally, in \cite{pavel2025hallucination} and \cite{zyberaj2025genai}, we show how LLM and Vision Language Model (VLM)-driven VSS interface specification extraction can be leveraged in order to reduce the occurrence of hallucinations in SDV code and tests. Finally, in \cite{zolfaghari2024rag} we adopt Retrieval Augmented Generation (RAG) based on Retrieve and Re-Rank methodology in order to enable LLM-based question answering relying on selected automotive standardization document. In this paper, we leverage a Retrieve and Re-rank pipeline based on \cite{petrovic2025ec} where a SentenceTransformer model (all-MiniLM-L6-v2) \cite{allMiniLM2024} encodes scenario queries and VSS/CAN entries to retrieve the top-k semantically similar signals based on embedding similarity. These candidates are then re-ranked using a cross-encoder (ms-marco-MiniLM-L-6-v2) \cite{msmarcoMiniLM2025} after which the highest-ranked entries are chunked to fit within LLM input constraints.

\begin{table}[t]
\caption{Summary of LLM-based solutions for automotive security}
\centering
\renewcommand{\arraystretch}{1.1}
\begin{tabular}{|p{0.17\columnwidth}|p{0.23\columnwidth}|p{0.45\columnwidth}|}
\hline
\textbf{Reference} & \textbf{Models} & \textbf{Use Case} \\
\hline

Nouri et al. \cite{nouri2025llm} &
CodeLlama, CodeGemma, DeepSeek-r1, DeepSeek-Coders, Mistral, GPT-4 &
Code co-generation and preliminary assessment pipeline for safety-related ADAS/AD systems. Evaluated on four automotive functions, including Adaptive Cruise Control (ACC) and Collision Avoidance by Evasive Manoeuvre (CAEM) \\
\hline

Khule et al. \cite{khule2025staf} &
GPT-4.1, DeepSeek-V3 &
Automated generation of executable security test cases (Python scripts) and Linear Temporal Logic (LTL) properties from attack trees for automotive security testing using the Security Test Automation Framework (STAF). \\
\hline

Louati et al. \cite{louati2025llmsecurity} &
OpenAI Codex (code-davinci-002) &
Automatic generation of formal security assertions (SystemVerilog Assertions) for autonomous vehicle hardware subsystems, including traffic signal controllers, AES encryption modules, and register access control mechanisms \\
\hline

Usman et al. \cite{usman2025darkside} &
HackerGPT (fine-tuned GPT-based LLM) &
Simulation and execution of cyberattacks on vehicle systems through automated generation of attack scripts targeting virtual CAN (vCAN), Bluetooth subsystems, and Remote Keyless Entry (RKE)/key fob systems. \\
\hline

\end{tabular}
\label{tab:llm_summary}
\end{table}

\section{Implementation overview}

\subsection{Code analysis workflow}
\begin{figure*}[htbp]
\centering
\includegraphics[width=\textwidth]{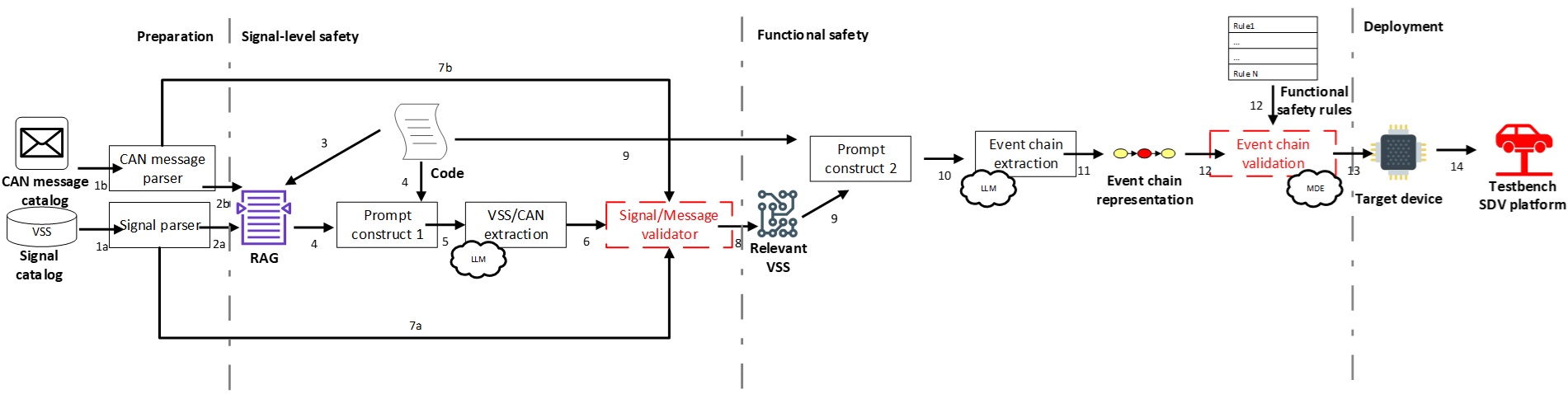}
\caption{Workflow of LLM-empowered event chain-based functional safety by design workflow for automotive.}
\label{fig:fs_analysis}
\end{figure*}

The proposed approach integrates Retrieval-Augmented LLMs with Model-Driven Engineering (MDE) to support secure and safe system design during early development of complex SDV system. The goal is to automate the extraction of relevant signals, reason about possible hazardous or insecure event chains, and validate them using formal design rules before deployment. Fig. \ref{fig:fs_analysis} shows the proposed workflow, where each step corresponds to one or more numbered element, as it will be described.

\textit{Inputs preparation (1a/1b)}: Vehicle signal catalog (VSS) and CAN message catalog (1a, 1b) are the authoritative, structured sources of truth (signal names, ranges, message IDs, semantics). They are forwarded to corresponding catalog parser which produce machine-readable signal/message representations containing only the relevant information, suitable for further steps of security and functional safety analysis. Those two catalogs are considered non-exclusively, as some of the programs executed within various parts of SDV system might rely on varying level of abstraction. In case of higher-level abstraction, common for SDV-targeting middleware (such as commAPI), VSS-based representation of signals as complex objects and corresponding handlers is often used, (1a). On the other side, when it comes to lower-level scripts, CAN messages might be used r referred to within the program directly as well (1b).

\textit{RAG-based context construction (2a/2b, 3)}: In practice, VSS catalogs usually contain thousands of entries, as well as hundreds of different CAN message types with additional variations might be used within the system, which would exceed LLM context and potentially lead to enormous token consumption, making the process costly. For this reason we make use of RAG layer, so it indexes the catalogs and prior artifacts so the further LLM-gnerated results are always grounded in concrete, up-to-date data, while minimizing hallucinations because RAG retrieves relevant catalog context with respect to the provided source code of SDV component configuration which is about to be analyzed by our workflow.

\textit{VSS/CAN message extraction (4-8)}: Prompt construct 1 aims to extract signals/messages from code/configuration.
VSS/CAN extraction produces a focused set of signals/messages relevant to the current security analysis. For this purpose, apart from the signal/message validator (a rule-based or formal checker) verifies that extracted signals/messages actually exist in the VSS/CAN catalogs and conform to expected formats and value ranges before they move forward. This prevents code and configurations with potentially malicious signal and message values to be run. The validated set of signals (Relevant VSS) becomes the input for event-level reasoning aiming further functional safety validation.

\begin{promptbox}[Prompt Construction 1 -- VSS/CAN message extraction.]
You are extracting list of VSS signals and CAN messages based on given source code \placeholder{code}. \\
For each of the steps signals/messages, extract entry: name, type, value, protocol.
 
\end{promptbox}

\textit{Event chain construction (9-11)}: On the other side, prompt construct 2 aims to generate event chain-based representation of the code, so the intended effect is represented as sequences of events, so insecure or unsafe outcomes can be easily identified. The LLM (with the RAG context) outputs structured event chains (e.g., “CAN message X forged → actuator command changed → vehicle state enters hazardous condition”). It builds upon our previous work from \cite{petrovic2025ec} Event chain extraction module converts the LLM text into a machine-readable event-chain representation, while in our case we rely on simple PlantUML notation of activity diagrams with some extensions. Based on our previous work \cite{petrovic2025meta}, it was shown that PlantUML \cite{plantuml} notation is less prone to errors while updated by LLMs compared to other solutions like Eclipse Modeling Frameworks's Ecore.
\begin{promptbox}[Prompt Construction 2 -- Event chain generation based on given code.]
You are updating PlantUml activity diagram about automotive event chain without comments and without explanations given as \placeholder{current-event-chain}, based on given source code: \placeholder{code}., taking into account \placeholder{relevant messages/signals}. \\
For each of event chain steps, the following parameters are considered as notes: input, input\_format, output, output\_format.
\end{promptbox}

\textit{Functional safety validation (11-12)}: Event chains are represented in the MDE environment as formal models (e.g., state machines, sequence diagrams, SysML/UML, or domain-specific models). In our case, we make use of PlantUML activtiy diagram which is converted to JSON format using model-to-model transformation for rule checking. Based on this JSON representation, we verify that the events appear in the correct order within the chain in order to enusre functional safety. For a rule specified as $e_1 \ \mathbf{before} \ e_2$, event $e_1$ must precede event $e_2$. The same applies inversely for rules using the after relation, where $e_1$ is required to occur after $e_2$.” Functional/ security rules (Rule1 … RuleN) exist as constraints, while the MDE tool validates the event chains against these rules. It is assumed that these rules are based on security references and good practices, defined either manually or relying also on LLMs (similar to our work from ). Example of rule usage would be to check that vehicle does not accelerate when camera detects pedestrian, so the rule "accelerate after pedestrian" should not be true within the examined event chain.

\textit{Deployment and testing (13-14)}: Verified artifacts (code for SDV modules, test scripts) are deployed to the Target device and executed in a Testbench / SDV (Software-Defined Vehicle) platform. This executes the hypothesized event chains and checks concrete safety/security consequences.

\textit{Feedback loops (7a/7b and 12 loop)}: If validation fails, feedback flows back to the user within the steps marked with red dashed line. This iterative loop closes the gap between LLM hypotheses and formally validated, testable artifacts. Corrective action can be taken either manually from user side, or relying also on LLM, considering the outcome of security analysis, as well as the code itself. In case of automated correction, we have another prompt constructed and executed against LLM, denoted as Prompt Construct 3.

\begin{promptbox}[Prompt Construct 2b -- Automated correction.]
Based on code analysis outcome \placeholder{result}, correct the following code \placeholder{code} to eliminate the detected functional safety-related issues. 
\end{promptbox}

\textit{Deployment Phase (step 13-14)}: The generated code is deployed to the target simulation environment (such as openly available digital.auto and CARLA) or embedded device (e.g., automotive ECU or zone controller) - either directly or over the air. 

\subsection{Security Analysis of System Topology}
The workflow integrates user inputs, metamodels, security guidelines, LLM-assisted model construction, and MDE-based design analysis to automatically determine whether the proposed automotive architecture satisfies communication-related security requirements. It builds upon our previous works from \cite{pan2025instance},\cite{pan2024ocl},\cite{petrovic2025meta}. This way, we provide synergistic, semi-automated automotive security analysis environment capable of validating complex system designs before proceeding further with implementation.

\begin{figure}[t]
    \centering
    \includegraphics[width=\columnwidth]{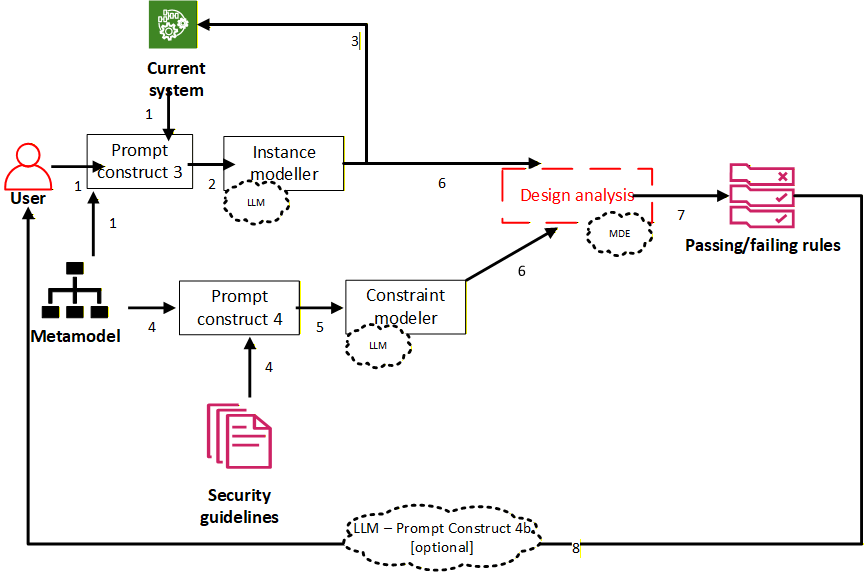}
    \caption{LLM-driven topology-level security by design workflow.}
    \label{fig:design_analysis}
\end{figure}

\textit{System Modeling Input: (Step 1)}: The user provides descriptions related to system topology - stem components, communication relationships, underlying protocols expected to be used for communication as well as other relevant architectural assumptions. Prompt Construct 3 is formed based on these inputs and transforms them into structured prompt suitable for an LLM.

\begin{promptbox}[Prompt Construct 3 -- System model creation/update.]
Update model instance \placeholder{current system}, with respect to \placeholder{metamodel}, based on requirements \placeholder{user input}. 
\end{promptbox}.
\textit{LLM-Assisted Instance Model Generation (Step 2)}: The LLM receives the structured prompt and generates an instance model, i.e., a concrete representation of the automotive system based on a predefined metamodel, including information about ECUs, sensors, buses, channels, protocols. This step builds upon \cite{pan2025instance}  and \cite{petrovic2025meta}. 

\textit{Current System feedback (Step 3)}: Current system representation using PlantUML class diagram is shown to the user, so additional features or refinements of the existing ones can be taken into account to update the model iteratively, going back to steps 1 and 2.

\textit{Security Guidelines Integration (Steps 3-5)}: Security guidelines in textual form (such as secure communication constraints, message format, allowed protocol usage between specific vehicle's components) are transformed to formal rules in OCL format with respect to the given metamodel, leveraging Prompt Construct 4. This work is based on our previous work from \cite{pan2024ocl}.

\begin{promptbox}[Prompt Construct 4 -- System model constraints generation.]
Generate automotive system security constraints with respect to \placeholder{metamodel}, based on reference specification \placeholder{security guidelines}. 
\end{promptbox}.

\textit{Model-Driven Security-Aware Design Analysis (Step 6-8)}: Previously generated EMF-compliant system model is checked against the security guidelines represented as OCL rules. Before that, previously generated PlantUML model is transformed to EMF Ecore format, using model-to-model transformation, as described in \cite{petrovic2025meta}. As outcome of analysis, user is informed whether security flaws based on the provided design have been identified, together with emphasize of on the failing rules. These rules are machine-readable and human-interpretable, so user can take corrective action - updating the system architecture, which results with re-generating model and re-running analyses, until all communication-related security constraints are satisfied. Optionally, LLM can also suggest corrections to the system model instance based on outcome, as defined by Prompt Construct 4b.
\begin{promptbox}[Prompt Construct 4b -- System model correction.]
Update automotive system model with respect to \placeholder{metamodel}, based on current representation \placeholder{current system} and analysis outcome \placeholder{OCL pass/fail list}. 
\end{promptbox}.

\section{Case Study Overview}
    \subsection{Functional Safety Analysis}
    In order to show proof-of-concept for functional safety aspects, we consider several experimental scenarios based on ADAS system that aims to perform emergency brake when either camera or LIDAR detects pedestrian. Each of these scenarios is considered from vehicular decision-making module perspective and has functional flaws that should be detected by our proposed LLM-empowered framework.
    
    - Scenario 1: Vehicle accelerates instead of braking when pedestrian is detected.
    
    - Scenario 2: Vehicle stops when LIDAR detects pedestrian, while only camera is used for sensing.
    
    - Scenario 3: Vehicle stops before pedestrian is detected.
    
\begin{figure*}[htbp]
    \centering
    \includegraphics[width=\textwidth]{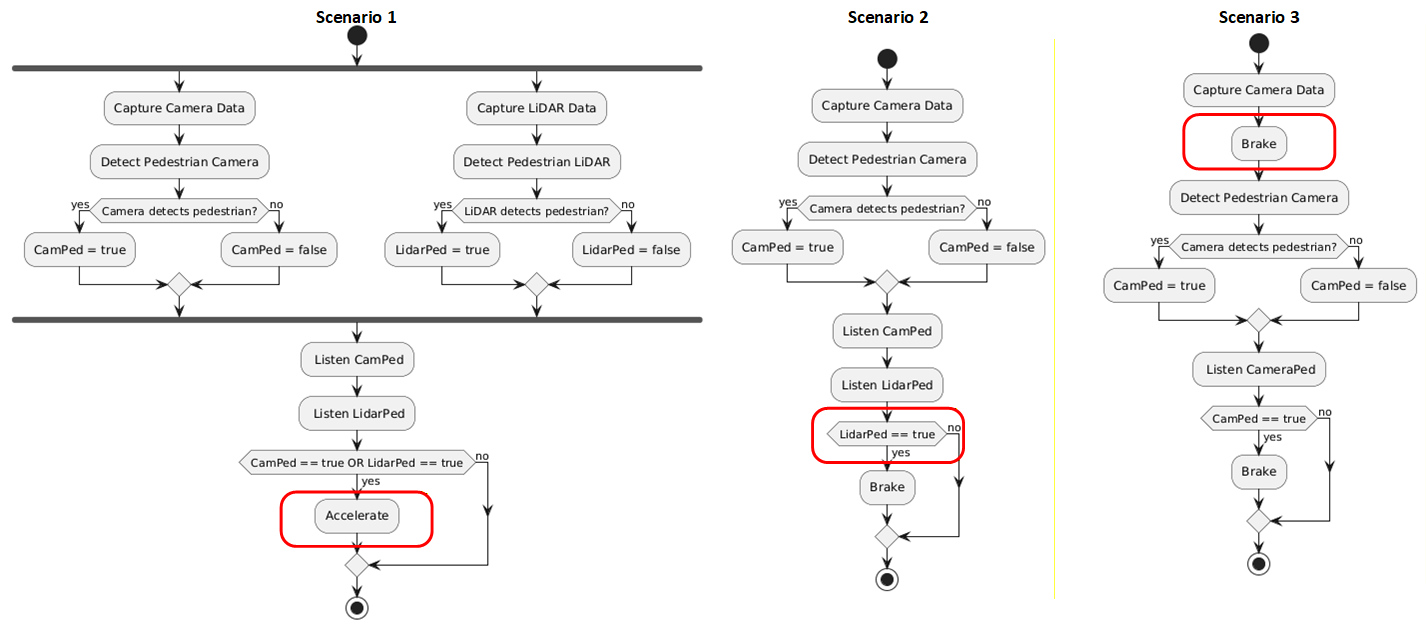}
    \caption{Overview of functional safety scenarios in activity diagram notation.}
    \label{fig:scenarios_sec}
    \end{figure*}

    In what follows, PlantUML-based activity diagram illustration of the given scenarios is shown in Fig. \ref{fig:scenarios_sec}. Critical parts of the diagram affecting functional safety are marked red for each of the scenarios. It is important to emphasize that for illustrative purposes and clarity, we depict simplified representations on conceptual level that do not contain directly exact CAN-FD messages or VSS signal full names within the diagrams themselves.
    
    In the first scenario, the risk is that vehicle will hit the pedestrian, as it will not stop when person is detected, but accelerate instead.
    For the second scenario, vehicle makes braking decision based on LiDAR information from the rest of the system, while only camera-based detection really exists within the system. This way, vehicle will likely not stop if pedestrian really occurs.
    In the third scenario, braking occurs before the decision-making, leading to vehicle early stopping.
    
    The underlying rules which are checked against the event-chain for each of the scenario are given in Table \ref{tab_rules_uc}.

    \begin{table}[t]
    \caption{Functional Safety Scenario Examples}
    \centering
    \renewcommand{\arraystretch}{1.1}
    \begin{tabular}{|p{0.1\columnwidth}|p{0.36\columnwidth}|p{0.38\columnwidth}|}
    \hline
    \textbf{Scenario} & \textbf{Rules} & \textbf{Description} \\
    \hline
    
    1 & 
    \textit{accelerate before pedestrian-camera-detected and accelerate before pedestrian-lidar-detected} & Acceleration command should occur only before the pedestrian is detected by either LiDAR or camera. \\
    \hline
    
    2 &
    \textit{(brake after camera-pedestrian detected and camera-sense before camera-pedestrian-detected) or (brake after lidar-pedestrian detected and lidar-sense before lidar-pedestrian-detected)} & Braking happens either as outcome of camera or LiDAR-based detection, while for each of them it is required that both sensing and analysis are done before triggering of actuator. \\
    \hline
    
    3 &
    \textit{brake after camera-pedestrian detected}& Braking happens only after pedestrian is detected by camera. \\
    \hline
    
    \end{tabular}
    \label{tab_rules_uc}
    \end{table}

\subsection{Topology Analysis}
\begin{figure*}[htbp]
    \centering
    \includegraphics[width=\textwidth]{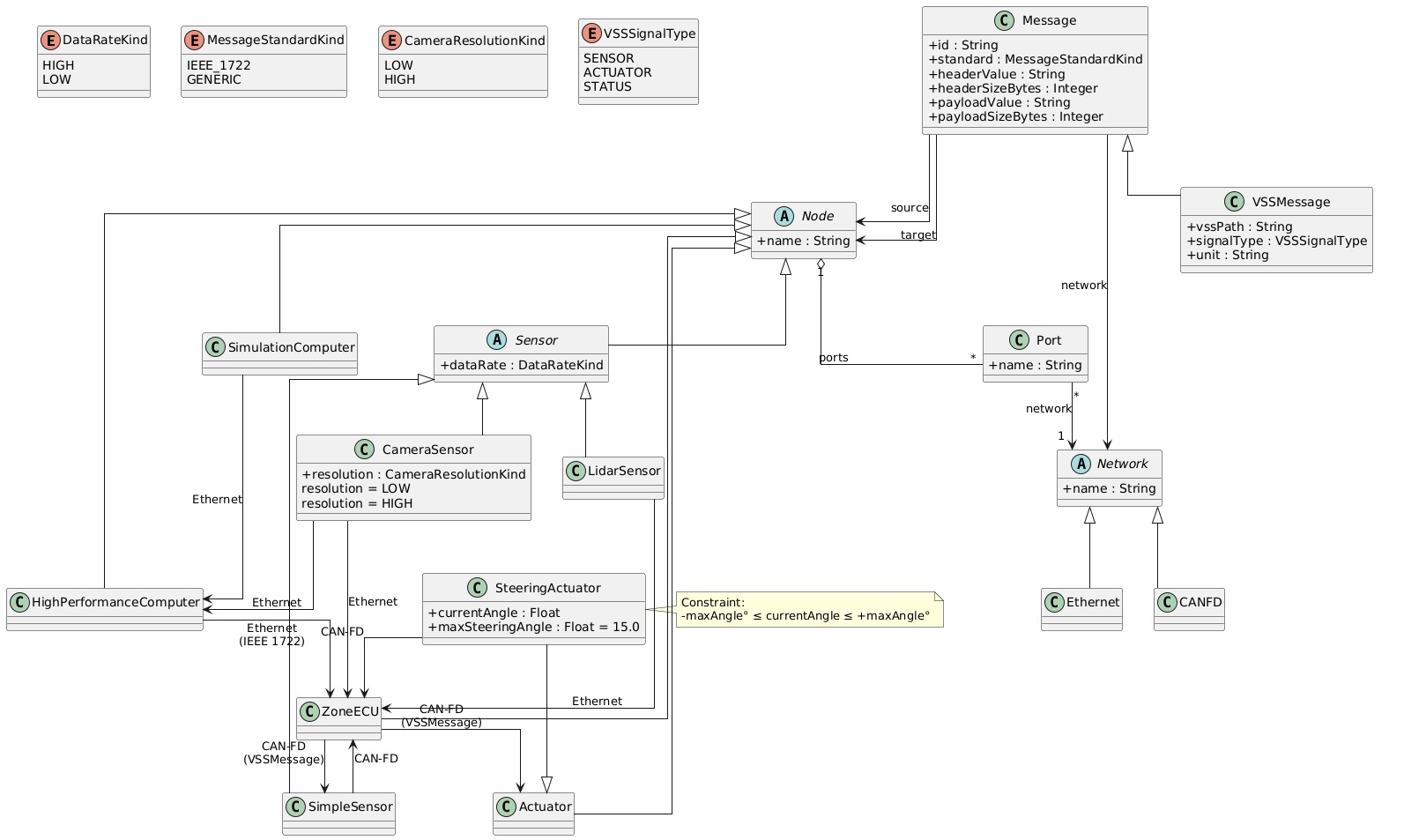}
    \caption{Metamodel behind the security-aware topology analysis case study.}
    \label{fig:sec_topology}
    \end{figure*}

Let us assume that user specifies component-level connectivity requirements: "Simulation computer is connected to in-vehicle high-performance computer via Ethernet. Moreover, in-vehicle High-performance computer is connected to Zone ECU via Ethernet and sends IEEE 1722-compliant messages to it. On the other side, high data rate sensors (such as lower resolution camera, LIDAR) are connected via Ethernet to Zone ECU, while simpler sensors and actuators are connected via CAN-FD. Higher resolution back and front cameras are connected to high-performance computer via Ethernet as well. Additionally, VSS message-level communication should be available from ZoneECU to connected devices via CAN, while it should include status aspects, actuator commands as well as sensing information".
Underlying metamodel is given in Fig. \ref{fig:sec_topology}. Additionally, example OCL rules encapsulating the main constraints when it comes to design-level security are given in what follows. The \emph{SteeringCommandWithinLimits} rule ensures that any message sent to a steering actuator contains a steering angle command within the safe range of -15 to +15 degrees, preventing unsafe steering commands at the communication level, as the physical testbench might be potentially damaged by larger angles.The \emph{HPCtoZoneEthernetIEEE1722} rule enforces the architectural constraint that all messages from the High-Performance Computer to the Zone ECU must be transmitted over Ethernet and comply with the IEEE 1722 standard, ensuring correct use of the backbone network. The \emph{TargetSpeedWithinSafetyLimit} rule guarantees that VSS messages defining the target vehicle speed do not exceed 30 km/h, enforcing a global safety limit on speed commands regardless of the sender or network.

\begin{promptbox}[Topology-level security analysis OCL rules.]
context Message \\
inv SteeringCommandWithinLimits:\\
  self.target.oclIsTypeOf(SteeringActuator) implies
    let angle : Real = self.payloadValue.toReal() in
      angle >= -15.0 and angle <= 15.0\\

context Message\\
inv HPCtoZoneEthernetIEEE1722:\\
  self.source.oclIsTypeOf(HighPerformanceComputer) and\\
  self.target.oclIsTypeOf(ZoneECU)\\
  implies
    self.network.oclIsTypeOf(Ethernet) and\\
    self.standard = MessageStandardKind::IEEE-1722\\

context VSSMessage\\
inv TargetSpeedWithinSafetyLimit:\\
  self.vssPath = 'Vehicle.Speed.Target'\\
  implies
    sSelf.payloadValue.toReal() <= 30.0

\end{promptbox}.

\section{Experiments and Evaluation}
Considering the fact that SDV-related requirements and source code are often subject to strict intellectual property protection regulations and constraints, adoption of locally deployable LLM-based solutions can be highly beneficial. In this paper, regarding the approach evaluation we compare effectiveness of two widely adopted models: commercial OpenAI's GPT-5 \cite{openai2025gpt5} against locally deployable Meta's Llama-3.3-70B-Instruct \cite{meta2024llama3-70b-instruct}. Taking into account our previous works which tackle with requirements handling and modelling, as well as assuming fixed RAG-based system for question answering, the experiments presented in this paper will focus on complementary aspects crucial to functional safety and security analysis not covered in our past publications - VSS/CAN signal mapping based on given software artifacts, as well as event chain-based functional safety validation. The evaluated LLM-empowered, integrated toolchain is also based on n8n\cite{n8n2025}, similar to our works from \cite{petrovic2025genai} and \cite{petrovic2025ec}. 

When it comes to the analyzed source code, we cover two execution environments based on Python: 1) testbench integrated with graphical CARLA simulator for autonomous driving \cite{wu2025viltum} and 2) digital.auto SDV simulation platform \cite{digitalauto}. There are three categories of experiments covered in Table \ref{tab_llm_evaluation_uc}: 1) CAN-FD mapping - identification of CAN messages and their values (payload, header) from the source code based on RAG-produced shortlisted candidate signals 2) VSS mapping - similar as 1), but in case of VSS signal objects 3) event chain-based functional safety - construction of event chain which can be further used for analysis based on Python source code, taking into account previously illustrated scenarios from Fig. \ref{fig:scenarios_sec}. For 1) and 2) we further consider two variants of the experiment - with smaller and larger number of assumed shortlisted signals produced by RAG-based system, while the RAG system itself is not evaluated in this paper.
The first column of Table \ref{tab_llm_evaluation_uc} denotes the experiment category, as previously described. The second column gives observations regarding the results observed while using GPT-5 model. Moreover, the third column shows per-experiment observations for locally deployable LLaMA model. The last column summarizes the model performance for specific experiment, expressed as percentage of successful completions in 10 runs for same conditions. In case of VSS and CAN-FD mapping, completion is assumed to be successful in case that all the correct signals are identified with their corresponding values. In case of event chain analysis, successful completion assumes that the constructed event chain was syntactically and semantically correct, so the functional safety flaws were identified based on the pre-defined rules for each of the scenarios, denoted as S1-S3. For both models, token limit per prompt was limited to 4096, while default temperature value was used. For execution of locally deployable mode, we rely on NVIDIA A100 GPU with 80GB of VRAM.

\begin{table}[t]
\caption{Evaluation of LLMs for Functional Safety Scenario}
\centering
\renewcommand{\arraystretch}{1.2}
\begin{tabular}{|p{0.21\columnwidth}|p{0.23\columnwidth}|p{0.22\columnwidth}|p{0.15\columnwidth}|}
\hline
\textbf{Aspect} & \textbf{GPT-5} & \textbf{LLaMA 3.3 70B Instruct} & \textbf{Performance} \\
\hline

CAN-FD mapping &
Correctly identifies CAN-FD signals and maps actuator and sensor events to message:value pairs with minimal ambiguity. &
Identifies most CAN-FD signals correctly, but occasionally produces redundant or loosely scoped message:value mappings. &
GPT (20): 90\% \newline
LLaMA (20): 80\% \newline
GPT (50): 90\% \newline
LLaMA (50): 70\% \\
\hline

VSS mapping (digital.auto) &
Accurately maps perception, sensing, and actuation events to standardized VSS paths (e.g., \texttt{Vehicle.Speed}, \texttt{ADAS.Brake}). &
Provides partial VSS mappings; some domain-specific signals require manual correction or normalization. &
GPT (20): 100\% \newline
LLaMA (20): 100\% \newline
GPT (50): 90\% \newline
LLaMA (50): 90\% \\
\hline

Event chain analysis (digital.auto) &
Consistently reconstructs correct causal chains (sense $\rightarrow$ detect $\rightarrow$ decide $\rightarrow$ actuate) aligned with functional safety rules. &
Captures main event orderings but may miss preconditions in complex rules. &
GPT: S1-90\%, S2-90\%, S3-100\% \newline
Llama: S1-70\%, S2-80\%, S3-80\% \\
\hline

\end{tabular}
\label{tab_llm_evaluation_uc}
\end{table}


\section{Conclusion}
This paper presents approach for SDV security and functional safety analysis of Python code leveraging LLMs in synergy with MDE, evaluated for basic ADAS capabilities in case of commercial and locally deployable model. Based on the achieved results, it can be noticed that both models have similar performance for VSS signal extraction, while commercial one is slightly better for CAN messages. It can be explained taking into account the fact that VSS notation is simpler for interpretation due to structured format and explicit signal hierarchy, while more expressive model is required to take into account implicit interpretation of CAN-FD format messages. Performance drops are observed in cases when larger lists of candidate signals are considered, while GPT performs slightly better in that case. However, the final results in this case are still mostly affected by the list of the selected signal by generated by RAG system.
On the other side, for event-chain based code analysis commercial solution is slightly better, while there are no significant differences regarding token consumption for both models. 

According to the outcomes, larger locally deployable models like the one evaluated in this paper are able to support SDV security and safety-related tasks with satisfiable performance even without fine tuning, assuming that contextual information (such as relevant signals and messages) is provided. Therefore, domination of LLM-based tools for security and functional safety in automotive is expected, due to increasing power of locally deployable models which enable to limit the flow of potentially confidential and protected software assets.

In future, our plan is to extend the adoption of the proposed solution to Rust programming language SDV ecosystem, which is becoming a promising direction in this area. As it ensures memory safety by design, we would aim to compare overall effectiveness and complexity of security and safety analysis of such approach to C++ and Python, as it promises simplification of such activities based on programming language-related design decisions. Additionally, we also aim to perform more detailed evaluation of the supporting RAG system alternatives, various parameter configurations and impact of locally deployable model fine-tuning.


\vspace{12pt}

\end{document}